\documentclass[10pt,letterpaper,twocolumn]{article} 
\usepackage{ol2}
\usepackage{amsmath}
\usepackage{hyperref}

\begin{document}
\twocolumn[ 
\title{Angular acceleration with radial dependence of twisted light}
\author{Jason Webster, Carmelo Rosales-Guzm\'an$^*$ and Andrew Forbes}
\address{School of Physics, University of the Witwatersrand, Private Bag 3, Wits 2050, South Africa\\
$^*$Corresponding author: \href{mailto:carmelo.rosalesguzman@wits.ac.za}{carmelo.rosalesguzman@wits.ac.za}}

\begin{abstract}
While photons travel in a straight line at constant velocity in free-space, the intensity profile of structured light may be tailored for acceleration in any degree of freedom.  Here we propose a simple approach to control the angular acceleration of light.  Using Laguerre-Gaussian modes as our twisted beams carrying orbital angular momentum, we show that superpositions of opposite handedness result in a radially dependent angular acceleration as they pass through a focus (waist plane).  Due to conservation of orbital angular momentum we find that propagation dynamics are complex despite the free-space medium: the outer part of the beam (rings) rotates in an opposite direction to the inner part (petals), and while the outer part accelerates so the inner part decelerates.  We outline the concepts theoretically and confirm them experimentally. Such exotic structured light beams are topical due to their many applications, e.g., optical trapping and tweezing, metrology and fundamental studies in optics. 
\end{abstract}
]

It is well-known that light propagates in free space at the constant speed $c$, a principle that lies at the foundations of relativity. Less well-known is the fact that this principle applies only to plane waves, light beams with a transverse spatial phase structure, known as structured fields, propagate at lower speeds \cite{Giovannini2015}. While structured light in general has found a multitude of applications \cite{Roadmap2017}, accelerating structured fields in particular have gained a great deal of interest since the first demonstration of freely accelerating light fields, the Airy beam \cite{Siviloglou2007a,Siviloglou2007}. Airy beams are solution to the paraxial wave equation in cartesian coordinates, which tend to accelerate along parabolic trajectories during propagation. Analogous solutions have been also demonstrated in different coordinate systems, as parabolic, oblate or prolate \cite{Bandres2008,Bandres2013} complemented by exact solutions to the Helmholtz equation \cite{Zhang2012}. The zoo of transversally accelerating beams extends to arbitrary convex trajectories and shapes \cite{Greenfield2011,Ruelas2014} and more recently to a new class provided also with radial acceleration, capable to evolve on parabolic spiraling trajectories \cite{Vetter2014}. Transversally accelerating beams have found many applications, e.g.,  to study fundamental aspects such as the interaction between optical vortices, \cite{Rosales2013}, to trap and guide particles along parabolic trajectories \cite{Baumgartl2008}, to enhance the image resolution in optical microscopy \cite{Vettenburg2014}, among others.

Of particular interest to us is the case of optical fields with angular acceleration. That is, fields that upon propagation tend to rotate about its optical axis while acquiring an angular acceleration. The simplest case of constant angular velocity has been extensively reported in the literature at both theoretical and experimental levels\cite{Tervo2001,Cerda1996,Allen1999,Rop2012a, Rop2012,Dudley2012}. Rotating light beams have been generated directly from a laser cavity \cite{Abramochkin1997}, through the use of customized diffractive optics \cite{Paakkonen198} or by superposition of beams carrying Orbital Angular Momentum (OAM) \cite{Schechner1996,Bekshaev2005,Kotlyar2007}, while angular acceleration was achieved via a superpositions of high order Bessel beams with non-linear azimuthal phases \cite{Schulze2015}.  Unfortunately such non-linear phases are not solutions to the paraxial wave equation and so had to be engineered via non-canonical superpositions of linear azimuthal phases.

In this letter we demonstrate, theoretically and experimentally, a novel method to control the angular acceleration of OAM carrying optical fields. For this we use the Laguerre-Gaussian ($LG_p^\ell$) modes as our OAM basis. Here, $\ell$ is an integer number, the azimuthal index, that accounts for the amount of OAM and $p\geq0$ is the radial index that accounts for the number of rings of maximum intensity. The ability to control both indices in $LG_p^\ell$ modes has proven to be very advantageous in optical communications, where the use of two degrees of freedom can potentially increase the number of information channels \cite{Trichili2016}. The superposition of two $LG_p^\ell$ modes, carrying opposite $\ell$ values, also generate a stationary rotation of the petal-like intensity distribution \cite{Bekshaev2005}. To induce acceleration, we exploit the Rayleigh length parameter $z_R$, engineering each mode with a slightly different $z_R$ value. The acceleration obtained in this way is radially dependent, an effect that causes the most outward ring to always rotates opposite to the inner ones. An analysis of such behavior was carried out to demonstrate that this is due to the need for conservation of orbital angular momentum. This type of rotating light fields could find applications in optical trapping and guiding of particles, in optical remote sensing as well as in microfluidic pumping \cite{Curtis2003,Arnold2007,Padgett2011,Lee2010,Shanblatt2011,Rosales2014}.

To construct accelerated light fields based on $LG_p^\ell$ modes, we superimpose two such modes with opposite helicity and slightly different Rayleigh lengths, denoted by $LG_p^\ell(z_{R1})$ and $LG_p^{-\ell}(z_{R2})$ as
\begin{equation}
u^{\ell}_p=LG_p^\ell(z_{R1})+LG_p^{-\ell}(z_{R2}).
\label{eq:ulp}
\end{equation}
We wish to stress that the beams given by $LG_p^\ell(z_{R1})$ and $LG_p^{-\ell}(z_{R2})$ above are on-axis beams with the same point of focus, same wave number, and that $z_{R1}$ and $z_{R2}$ only differ enough to allow the two beams to interfere over a desirable propagation length. The purpose of adding beams with opposite OAM is to reveal the underlying structure of the Laguerre-Gauss modes through interference effects, which will provide a means for us to track the acceleration of the field.

If we denote $I_1$ and $I_2$ as the transverse intensity of the $LG_p^\ell(z_{R1})$ and $LG_p^{-\ell}(z_{R2})$ beams respectively then the resulting intensity $I$ of the $u^{\ell}_p$ field will be given by
\begin{equation}
I=I_1+I_2+2\sqrt{I_1 I_2}\cos\left[2\left(\ell\phi+\Delta R_{12}(r,z)-\Delta \psi_{12}(z)\right)\right],
\label{eq:int}
\end{equation}
where,
\begin{align}
\Delta \psi_{12}(z)&=\frac{\psi_1(z)-\psi_2(z)}{2}, \\
\Delta R_{12}(r,z)&=\frac{k r^2}{4}\left(\frac{1}{R_1(z)}-\frac{1}{R_2(z)}\right),
\end{align}
$R_1(z)$ and $R_2(z)$ denote the radius of curvature of the beams with Rayleigh length $z_{R1}$ and $z_{R2}$, while $\psi_1(z)$ and $\psi_2(z)$ are the Gouy phase of each beam respectively. Finally, throughout the rest of the paper, we will use $z_R$ as the average value between $z_{R1}$ and $z_{R2}$.

\begin{figure}[htbp]
	\centering
	\includegraphics[width=\linewidth]{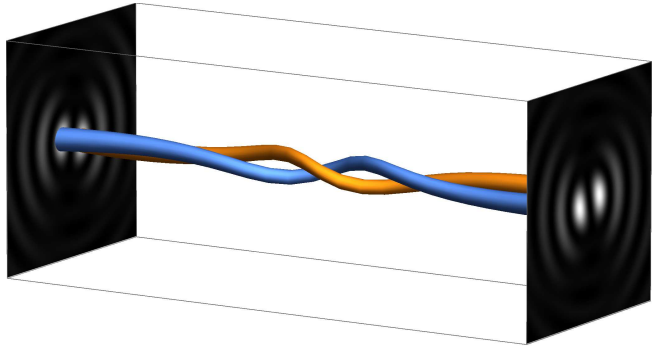}
	\caption{Tridimensional representation of the inner petals rotation during propagation in free space. To highlight the rotation we have used an $\ell=1$, $p=10$ beam for this figure, focusing only on the inner petal structures.}
	\label{fig:3D}
\end{figure}

In order to track an specific point of the field during propagation, we select stationary points in the intensity profile to follow.   According to this, any point inside the beam rotates in a manner described by
\begin{equation}
\Phi(r,z)=\frac{\Delta \psi_{12}(z)-\Delta R_{12}(r\text{ } w(z),z)}{\ell},
\label{eq:rot}
\end{equation}
where $w(z)$ is the beam width normalized such that it is unity at the focal point. This normalization is required since the generated beam focuses during propagation and thus any point of interest will also focus, leading to a change in its radial distance. In this case, $r$ is the measured radius of our desired point at the beam waist. This is a non-linear rotation, that is also radially dependent, and arises directly from the interference seen in equation \ref{eq:int}. This radial dependence was not seen in the angularly accelerating Bessel beam presented in \cite{Schulze2015} and as such presents a novel form of accelerating optical fields. In fact, this radial dependence causes the various rings in an LG mode to rotate at different rates and causes the outermost ring to rotate in a different direction to the inner petal structures. The rotation upon propagation of the inner most petals is graphically represented in Fig. \ref{fig:3D}.

The angular acceleration can be obtained by computing the second derivative in equation \ref{eq:rot} w.r.t. $z$. This produces
\begin{equation}
\Delta \ddot{\psi}_{12}(z)=z(1+2p+2|l|)\left[\frac{z_{R2}}{\left(z^2+z_{R2}^2\right)^2}-\frac{z_{R1}}{\left(z^2+z_{R1}^2\right)^2}\right], \\
\end{equation}
\begin{align}
\Delta \ddot{R}_{12}(r,z)=\frac{zk r^2}{2z_R^2}&\Bigg[\frac{\left(z_{R}^2-z_{R1}^2\right)\left(z^2-3z_{R1}^2\right)}{\left(z^2+z_{R1}^2\right)^3} \nonumber \\
&+\frac{\left(z_{R2}^2-z_{R}^2\right)\left(z^2-3z_{R2}^2\right)}{\left(z^2+z_{R2}^2\right)^3}\Bigg],
\end{align}
resulting in
\begin{equation}
\ddot{\Phi}(r,z)=\frac{\Delta \ddot{\psi}_{12}(z)-\Delta \ddot{R}_{12}(r\text{ } w(z),z)}{\ell}.
\label{eq:accel}
\end{equation}

To experimentally generate the accelerated optical fields, we used a 10 mW linearly polarized He-Ne laser ($\lambda=632.8$ nm). The beam was expanded and collimated to illuminate a HoloEye Pluto spatial light modulator (SLM) in which we encoded the required phase pattern to produce the desired $u^{\ell}_p$ field by complex amplitude modulation \cite{Forbes:16}. For the experiments we chose $z_{R1}=5.8926$ cm and $z_{R2}=3.0601$ cm. The intensity profile of the accelerated beam created in this way was tracked over a propagation range $z_{range}= 11.119$ cm using a CCD camera (Point Grey fire-wire) . The beam propagation was performed digitally, with no moving parts, using the angular spectrum approach programmed onto our SLM \cite{Schulze:12}, and observed at the Fraunhoffer plane through the use of a lens of focal length $f=150$ mm. We captured 230 images of the intensity profile at regular intervals throughout $z_{range}$, taking the focusing plane as the center.
\begin{figure}[htbp]
	\centering
	\includegraphics[width=\linewidth]{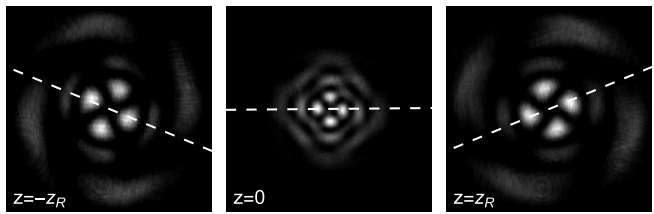}
	\caption{Animation of the experimentally generated intensity profile of free-space propagating LG modes with angular acceleration for the $\ell=2$, $p=2$ beam (see Visualization 1).}
	\label{fig:mov_rotation}
\end{figure}

\begin{figure*}[htbp]
	\centering
	\includegraphics[width=\linewidth]{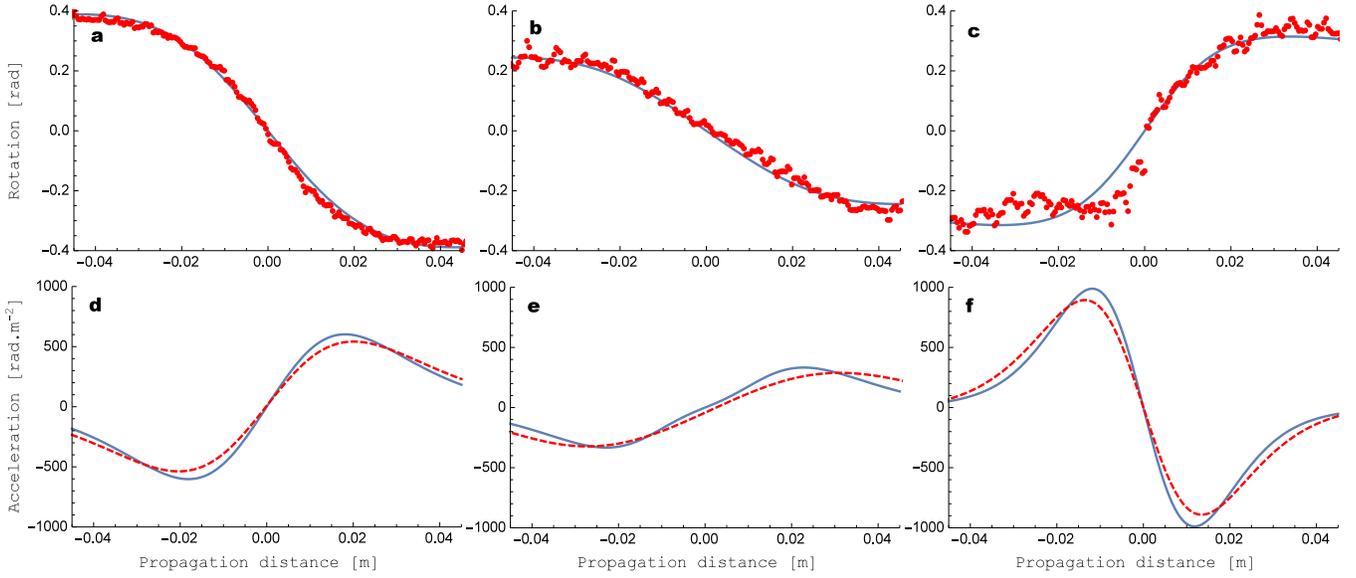}
	\caption{(Colour online) Measured (red) and theoretically predicted (blue) values for the rotation and acceleration of our $l=2$, $p=2$ accelerated field plotted against the propagation distance. The figures correspond to (a) inner petals ($r=99.7392$ $\mu$m), (b) first ring ($r=134.857$ $\mu$m), and (c) second ring ($r=222.856$ $\mu$m), with the same correspondence for the acceleration figures labeled (d)-(f).}
	\label{fig:rotandacc}
\end{figure*}

We generated a set of twelve individual $u^\ell_p$ fields in our experiment with indices ranging from $\ell=[1,3]$ and $p=[0,3]$, with an animation of the $\ell=2$ and $p=2$ example provided in Fig. \ref{fig:mov_rotation}. To measure the rotation we first removed the focusing effect of the acquired images. Then, reference images were chosen and masked to isolate regions of interest, such as the inner petal or outer ring structures. To measure the rotation, all the images taken along $z_{range}$ were correlated with a reference image, and the radial dependence in our rotation was measured by changing the structure of interest in our reference image. The average error in these experimental measurements was found to be roughly $0.5^\circ$. We've also set the rotation of the image at the beam waist to be zero.

The resulting rotation of our accelerating $l=2$, $p=2$ field is shown in Figs. \ref{fig:rotandacc}(a)-(c), plotted with the theoretical prediction for the first, second, and third rings respectively. The theoretical predications are more approximations in this case, as we are tracking entire structures in our measurements, not individual points. Notice the good agreement between the experimental and the theoretical predictions. The most interesting feature from this rotating field is that the outermost ring rotates in an opposite direction to the inner structures, as seen by noting the opposite sign of the rotation in Figs. \ref{fig:rotandacc} (a) and (c).

To quantify the angular acceleration we used an interpolating function approach to find the derivatives of the rotation data so as to reduced the effects of random noise (which introduces artifacts in the results when taking derivatives) The calculated angular acceleration curves from the experimental data is seen in Figs \ref{fig:rotandacc}(d)-(e) together with the theoretical curves. 

\begin{figure}[htbp]
	\centering
	\includegraphics[width=\linewidth]{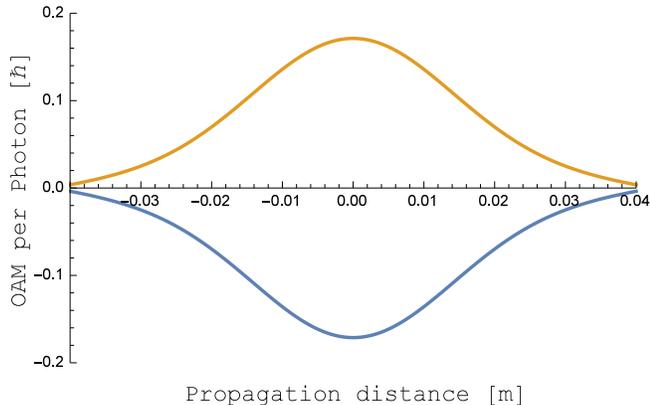}
	\caption{(Colour online) The orbital angular momentum per photon of the inner (blue) and the outer (orange) field structures, defined by $L_{inner}$ and $L_{outer}$ respectively, plotted against the propagation distance for our $l=2$, $p=2$ field that we have been studying. The total OAM is thus precisely zero, leading to a conserved OAM throughout the propagation.  This accelerating and decelerating behaviour can be seen in Figs \ref{fig:rotandacc}(d)-(e).}
	\label{fig:ang}
\end{figure}

As previously mentioned, an interesting feature of our rotation is the fact that the outermost rings rotate in a direction opposite to that of the inner petals. This phenomenon is required for the conservation of angular momentum within the field. To prove this, we used the orbital angular momentum density given by \cite{Litvin:12}
\begin{equation}
L_z=\frac{1}{c^2}\left[\mathbf{r}\times\mathbf{S}\right]_z,
\end{equation}
where $\mathbf{S}$ is the Poynting vector
\begin{equation}
\mathbf{S}=\frac{\varepsilon_0 \omega c^2}{4}\left[i\left(u\nabla u^*-u^*\nabla u\right)+2k|u|^2\hat{z}\right]. \\
\end{equation}
To probe that the total angular momentum is conserved during propagation, we will show that the angular momentum of the inner rotating field precisely cancels with that of the outer counter-rotating field. The distinction between these two fields can be found at a radial point $r_0$ by setting Equation \ref{eq:rot} to zero and solving for $r_0=r w(z)$ (to remove the tracking effect). The result yields only one rotation crossover point, namely
\begin{equation}
r_0(z)=\left[\frac{4 \Delta \psi_{12}(z)}{k}\left(\frac{1}{R_1(z)}-\frac{1}{R_2(z)}\right)^{-1}\right]^{1/2},
\end{equation}
which is a function of $z$ alone and focuses much in the way we would expect it to. The inner rotating structures always lie within $r<r_0$ while the outermost counter-rotating ring lies at $r>r_0$. We can now define the inner and outer total orbital angular momentum quantities as
\begin{align}
L_{inner}(z)&=2\pi\int_0^{r_0(z)} L_z r dr \\
L_{outer}(z)&=2\pi\int_{r_0(z)}^{\infty} L_z r dr
\end{align}
If we plot these quantities (per photon) against $z$ then we see that $L_{inner}$ precisely cancels with $L_{outer}$ as shown in Fig. \ref{fig:ang}. Thus the total orbital angular momentum $L_z=L_{inner}+L_{outer}$ is zero and is conserved throughout the propagation of the beam.  This is a noteworthy result as angular acceleration by definition implies that the angular momentum of the beam must be changing, which is forbidden for free-space propagation.  Now we see that the radial dependence of the acceleration is a fundamental requirement for the conservation of OAM: as one part of the beam gains OAM through acceleration, so the other regions lose OAM through deceleration.

In conclusion, we have outlined and demonstrated a simple method to generate twisted light fields with angular accelerations. This method is based on the superposition of two Laguerre-Gaussian modes with opposite helicity and differing Rayleigh lengths. This way of generating rotating light fields causes a radial dependence on the angular acceleration. Of particular interest is the fact that the outer part of the beam rotates in an opposite direction to the inner part, the former accelerating while the later decelerating. We analyzed this behavior in terms of the angular momentum density and realize that while one part gains angular momentum, the other loses, leading to  a conservation in the total angular momentum within the optical field, despite the angular acceleration.

\end{document}